\documentstyle[psfig,12pt]{article}
\author{Z. Kozio\l \thanks{electronic addresses:
zkoziol@is.dal.ca; http://is.dal.ca/$\sim$ zkoziol/zkoziol.html}  \space and R.
A. Dunlap \\
{\it Department of Physics, Dalhousie University} \\
{\it  Halifax, N.S., Canada B3H 3J5}}
\title{{\bf Nonlinear Flux Diffusion and \\
ac Susceptibility of Superconductors \\
- Exact Numerical Results}}
\date{August 18, 1995}
\begin{document}
\maketitle
\begin{abstract}
The ac response of a slab of  material with electrodynamic
characteristics $E\sim j^{\kappa+1}$, $\kappa\geq0$,
is studied numerically. From the
solutions of the nonlinear diffusion equation, the
fundamental
and higher-order components of the harmonic  susceptibility
are
obtained. A large portion of the data for every $\kappa$ can
be scaled by
a single parameter, $\xi$
=$t^{1/(\kappa+2)}\cdot H_0^{\kappa/(\kappa+2)}/D$,
where $t$ is the period of
the ac field at the surface, $H_0$ is its amplitude and $D$
is the slab
thickness. This is, however, only an approximate scaling
property: The field penetration into a nonlinear medium is a
more
complex phenomenon than in the linear case. In particular,
the
susceptibility values are not uniquely defined by a set of
only
two parameters, such as $\kappa$ and $\xi$, while one
parameter, i.e. $t^{1/2}$/D,
is sufficient to describe the electrodynamic response of a
linear
medium.
\end{abstract}

\vspace{5mm}
\newpage

   {\bf 1. Introduction: Nonlinear diffusion and ac
susceptibility of superconductors.}

   \ The problem of nonlinear diffusion has recently
attracted
considerable attention in diverse fields of science. One
example,
which is considered in the present work, deals with the
magnetization process of superconductors. In the case when
the
response of a superconductor to an applied ac field,
$H=H_0\cdot exp(\it{i} \omega t)$,
is linear, the electrodynamic properties in the
superconducting
state can be described in terms of the complex conductivity,
$\sigma =\sigma _r+\it{i} \sigma _i$. In the flux--flow
regime of high--$T_c$ superconductors, which
occurs in a broad $H-T$ range, it is usually justified to
neglect the
imaginary component of the complex conductivity. In that
case,
Maxwell's equations, $\nabla\times{\bf B}=4\pi/c\cdot {\bf
j}$
and $c\nabla\times {\bf E}=-\partial {\bf B}/\partial t$,
provide the
following result for the field penetration into a material
filling half- space, $x>0$:
$B=H_0\cdot \eta(x) \cdot exp(\it {i} \omega t)$,
with $\eta(x) = exp(- \lambda x)$, and
$\lambda ^2 = 2\it{i} / \delta ^2$,
where $\delta ^2=c^2 /( 2\pi \omega \sigma _r)$.
For a thin plate of thickness $2D$, with the ac
field parallel to the large surface of the plate,
the following equations hold [1]:

\begin{equation}
\label{e1}
4\pi \chi ' = -1+ \frac 1{a} \cdot
\frac {sinh(a)+sin(a)}{cosh(a)+cos(a)}, \\
\space 4\pi \chi ''  = \frac 1{a} \cdot
\frac{sin(a)-sinh(a)}{cosh(a)+cos(a)},
\end{equation}

where $a=2D/ \delta $. A maximum in $\chi ''$ results
when $\delta (\sigma _r, \omega)$ is comparable
to the sample size.
Experimental results rarely show susceptibility
curves which correspond to the flux-flow result of ohmic-
like
behaviour. Rather, the effects are most often nonlinear. The
dependence of the measured susceptibility curves on the
excitation
current and higher-order components in the harmonic
susceptibilities are observed. In the limiting case of very
strongly
nonlinear response, the critical-state model may be used for
the
calculation of the ac susceptibility. Then, only one
parameter is
needed to construct the hysteresis curve for the
magnetization, the
field of the first full penetration to the sample center,
$H^*$. It is
assumed that the harmonic susceptibility components,
$\chi ' _m$  and $\chi '' _m$ ,
are defined as Fourier components of the time-dependant
magnetic
hysteresis curve; $M(t)/D=\sum _m \left( \chi ' _m \cdot
cos(m \cdot \omega t) + \chi '' _m \cdot sin(m \cdot \omega
t) \right)$,
where $m$ is an integer. When, for instance,
$H_0<H^*$, $4\pi \chi ' _1 = - (1-H_0/2H^*)$ is
obtained then, $4\pi \chi ' _m = 0$ for every odd $m>1$,
and $4\pi \chi '' _m =2H_0/3\pi mH^*$ for all odd $m>0$.
To deal with situations which are
more relevant to the description of real experimental
results, it is
necessary to investigate a nonlinear theory of the magnetic
response which would bridge the two limiting cases observed:
the
linear-response and the critical-state one. A fruitful
approach to
this problem is based on studies of the electrodynamic
response of
a medium characterized by a power-law current-voltage
dependence, $j= \sigma (E) \cdot E= \sigma _0
\cdot E_0 \cdot (E/E_0) ^{1/(\kappa +1)}$ ,
where $\kappa \geq0$.  Using
Maxwell's equations, the nonlinear diffusion equation
describing
the penetration of fields into a slab of thickness $2D$
lying in the $yz$ plane [1-3] can be derived as

\begin{equation}
\label{e2}
 {\partial \beta \over \partial \overline{t}}=
 {\partial \over \partial \overline{x}}
  \left(
    {\partial \beta \over \partial \overline{x}} \cdot
{
    \left|
       {\partial \beta \over \partial \overline{x}}
    \right|
}           ^{\kappa}
  \right),
\overline{t}={t \over \tau _0 }, \overline{x}={x \over x _0
},
\end{equation}

where $ \beta =B/E_0$, $x_0 = c/(4 \pi \sigma _0)$
 and $\tau _0 = 1/(4 \pi \sigma _0)$. Recently, studies of
 solutions of eq. (2) have been carried out by many authors
[2-4].
The exact analytical description of  the response of a
superconductor to an abrupt change of external field has
been
given in references [1] and [2] and is compared with the
results of
non-logarithmic magnetization relaxation measurements on
high-$T_c$ materials [5]. Various aspects of the ac
response of superconductors has been studied as well by
Dorogovtsev [6]
and van der Beek et al. [7]. Recent results of Gilchrist and
Dombre [8] can be compared with numerical results described
in
the present work. The distinctive feature of solutions of
eq. (2)
is that the flux-profile penetration resembles that in the
models
of the critical-state; When a field change is applied, the
profile
of perturbation spreads out from the surface towards the
sample center but a region exists where the field
distribution is
unchanged inside. If the response to a field change of $H_0$
is
considered, the time after the front of the field change
arrives
to the center, $t^*$, is given by,

\begin{equation}
\label{e3}
{t^* \over \tau _0}=
{\kappa \over 2 (\kappa +1)}{1 \over (\kappa +2)} \left(
{\Gamma(1/\kappa +1)\Gamma (3/2) \over \Gamma (1/\kappa
+3/2)} \right) ^{\kappa} \cdot \left({D \over x_0} \right)
^{\kappa +2}.
\end{equation}

The initial magnetization, at $t<t^*$, is given by:

\begin{equation}
\label{e4}
4\pi M=-H_0 \cdot \left(1- \left({t \over t^*}
\right)^{1/(\kappa +2)} \cdot
{\Gamma (1/\kappa +3/2) \over \Gamma (1/\kappa +2) \cdot
\Gamma (1/2) } \right).
\end{equation}

 \ Equations (3) and (4) imply that a single parameter,
$\xi \equiv t^{1/(\kappa +2)} \cdot H_0 ^{\kappa /(\kappa
+2)}/D$,
can parametrize the short-time magnetization
relaxation. It is informative to determine the extent to
which
this scaling relation is valid with respect to the ac
susceptibility
 (with the replacement of $t$ and $H_0$ by the ac field
period and the
field amplitude, respectively).

 {\bf 2. Numerical modelling of nonlinear diffusion.}

   \ Most of the calculations of the nonlinear diffusion
process
presented here have been performed on an array of dimension
$50 \times 200$, containing magnetic induction values, $B$,
at $50$ time
intervals and $200$ space-intervals [9]. The magnetic field
at the
surfaces of the sample, $H_0 \cdot sin( \omega \cdot t)$,
determines the boundary conditions.
An average magnetic field $<B(t)>$ in the sample has
been computed from the magnetic field distribution,
every $50$ time steps. Next, a Fourier time-analysis of
$<B(t)>$ has been performed and the coefficients of the
fundamental and higher-
order terms of the harmonic content have been found,
$4\pi \chi ' _m =<B' _m>$ and $4\pi \chi '' _m = <B'' _m>$,
except for the real component of the fundamental
susceptibility,
which is given by $4\pi \chi ' _1 = -1 + <B' _1>$.
The method of computation and its results have been
carefully
tested. First, the magnetization relaxation process after an
abrupt change of the external field has been simulated and
numerical results were compared with the known exact
analytical epressions derived by Kozio\l \space and de
Chatel [2].
Then, the validity of the modelling of the ac-response in
the limit of
linear diffusion on the ac susceptibility, as given by eq.
(1), was
checked. It was confirmed also that the ac susceptibility
converges towards the critical-state results for large
$\kappa$. The time
range for which the response to the ac field becomes
periodic
(the initial response at short time does not satisfies this
condition) was also investigated. In most cases it is safe
to
analyze the data taken after the initial $50000$ steps in
time
evolution (this time depends on $\kappa$ and $H_0$).
An additional, more reliable, criterion of stable
periodicity is
based on the criterion that the dc or second harmonic
components
are not found. At
large values of $H_0$, the calculations become unstable
abruptly. It
is possible to overcome this difficulty but at significant
expense
in computation time (the computation of one susceptibility
point
requires an average of about 3 hours on an IBM-PC computer
486DX2-33MHz). Therefore, we have concentrated on
performing calulations for a larger number of points at
lower
fields.

\vspace{5mm}

{\bf 3. Results and discussion}

  \ The penetration of an alternating field resembles, in
some
ways, the response to an abrupt change of external field;
the
amplitude of field changes diminishes gradually in the
material
and, if the field amplitude at the surface is not too large,
there
is no penetration to a volume separated by a certain
distance
from the sample surface. Whether the front of the flux
profile in
ac penetration propagates towards the center or not, is not
an
easy question to answer, since the initial very slow
propagation
which is observed might only be due to unstable initial
conditions. Within the accuracy of calculations, the flux
profile
has a self-replicating shape of diminishing amplitude, with
perfect periodicity in time at every point in space but with
a
phase shift which changes with the distance from the
surface.
The profiles obtained for one value of an ac field amplitude
coincide with the profiles computed for another ac field
amplitude, if the phase lag and spatial coordinates are
shifted
properly.
	Plots of $\chi ''$ versus  $\chi '$  shown in Figures 1
and 2
for different values of the nonlinearity parameter $\kappa$
converge to the limit of linear diffusion for $\kappa
\rightarrow 0$
and to the limit given by the critical-state model for large
$\kappa >>1$.

\vspace{5mm}
\centerline{\psfig{figure=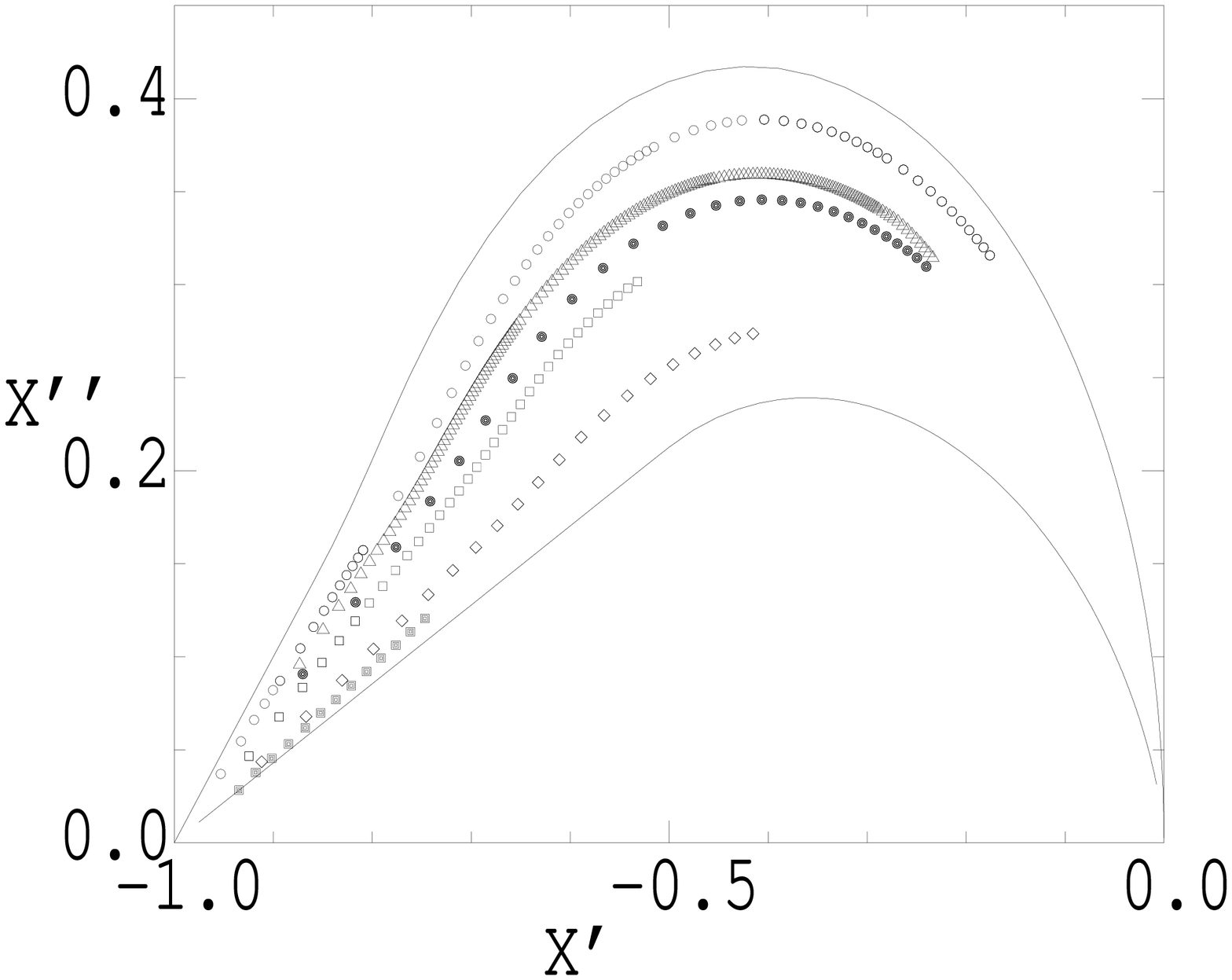,height=85mm,width=100mm,bbllx=0pt,bblly=0pt,bburx=600pt,bbury=650pt,angle=270}}
{\bf Figure 1.}
{\sl  The $\chi ''$ versus  $\chi '$ plots of the ac
susceptibility for different values of the nonlinearity
parameter $\kappa$ and different periods of the ac field
$t$, ($\kappa ,t$):
(0.667, 25000, $\circ$ ), (0.667, 5000, $\Delta$), (2,
12500, $\bullet$  ), (2, 5000, $\Box$ ), (3, 10000,
$\diamond$ ), (12, 6250, $\Box$ ). Solid lines represent the
critical-state and the
linear-response limits for a thin plate. The difference
between the data for $\kappa =2$ obtained for two different
frequencies of the ac field should be noted.}
\vspace{5mm}

\centerline{\psfig{figure=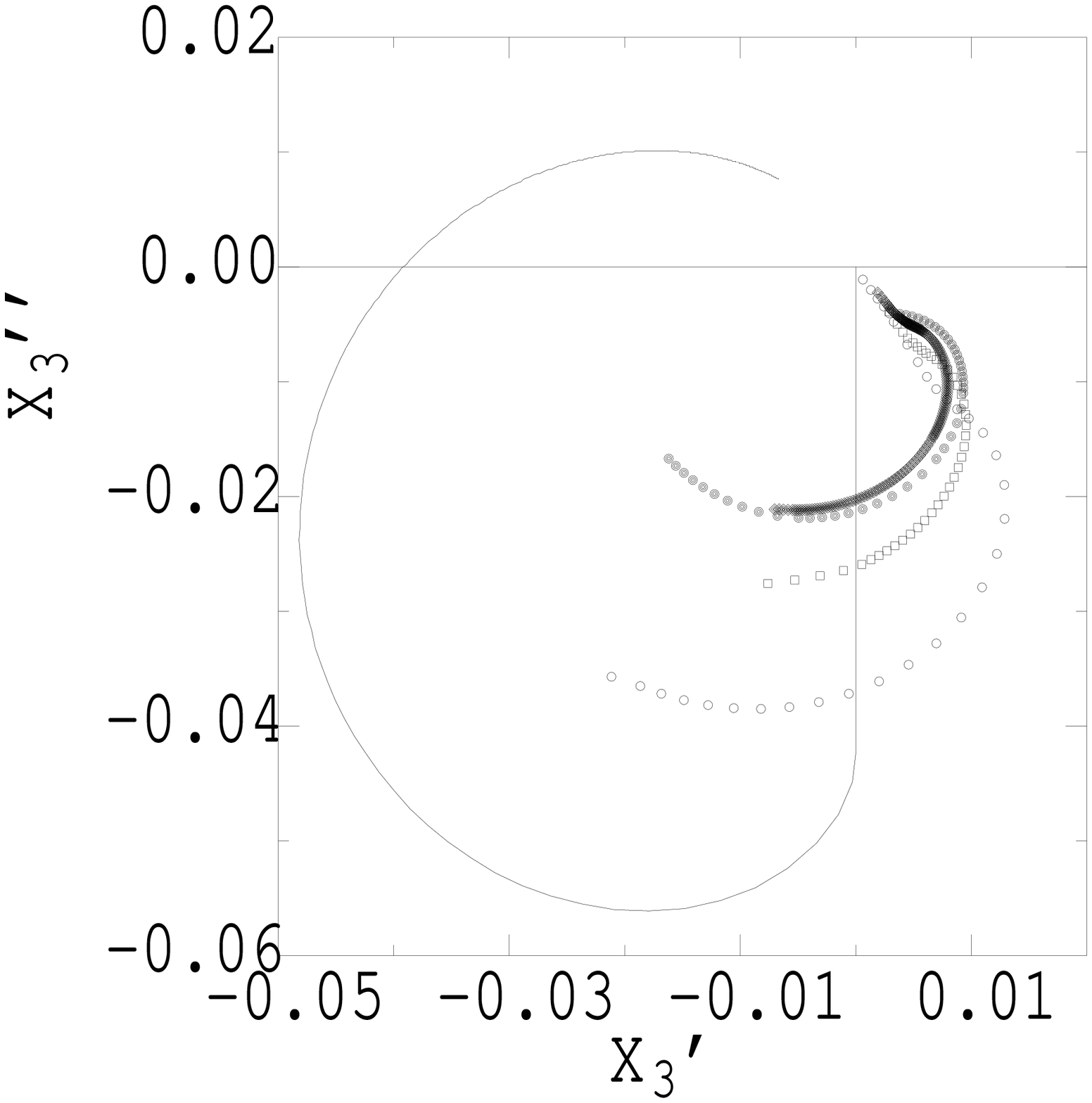,height=75mm,width=100mm,bbllx=0pt,bblly=0pt,bburx=600pt,bbury=650pt,angle=270}}
{\bf Figure 2.}
{\sl  The third harmonic susceptibility $\chi '' _3$ versus
$\chi ' _3$ compared to the critical-state result
represented by the solid line, for the following
nonlinearity parameter $\kappa$ and periods of the ac field
$t$, ($\kappa ,t$): (0.667, 25000, $\bullet$ ),   (0.667,
5000, $\Diamond$ ),  (1, 5000, $\Box$) and (2, 12500,
$\circ$).}
\vspace{5mm}

An important feature of the
present results is seen in the Figure 1; Susceptibility
points
computed for different frequencies of the ac field but the
same
value of $\kappa$, do not fall on the same curve. This is
different from
what it might be expected and seems to have been unnoticed
in
previous work [8]. In Figure 3, we show that a simple
scaling of
the susceptibility with the amplitude of the ac field holds
for the
data obtained in the range of incomplete flux penetration,
$\chi \sim H_0 ^{\kappa /(\kappa +2)}$.
In Figure 4, the real component of the first harmonic
susceptibility is drawn as a function of $H_0 ^{\kappa
/(\kappa +2)} \cdot
t ^{1/(\kappa +2)} /D$, for
different values of the field amplitude $H_0$, period of the
field and
for a few sample sizes. This latter scaling method is not
perfect;
small differences in the slopes of the data computed for
various
frequencies is found. This effect may be explained by the
fact
that flux profiles have a shape which depends on the time of
field penetration. One should expect that the parameter
$\xi =t ^{1/(\kappa +2)} \cdot H_0 ^{\kappa /(\kappa +2)}/D$
will become an exact scaling variable only for the
cases when all the parameters, $t$, $H_0$ and $D$, are
simultaneously
scaled by a constant $\lambda$ in the following way: $D
\rightarrow \lambda
\cdot D$, $H_0 \rightarrow (\lambda \cdot H_0) ^{\kappa
/(\kappa +2)}$
and $t \rightarrow (\lambda \cdot t)^{1/(\kappa +2)}$.

\vspace{5mm}

\centerline{\psfig{figure=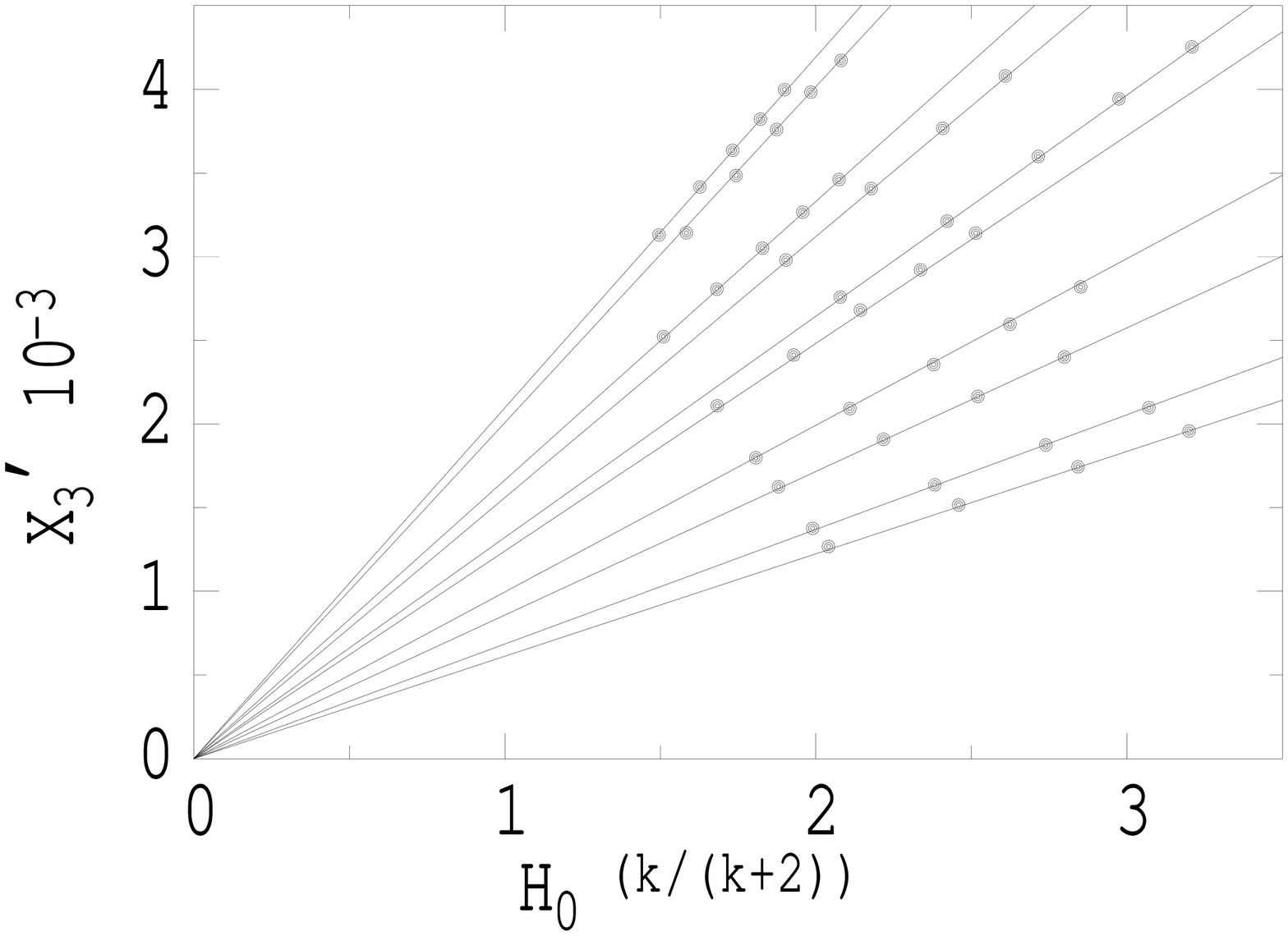,height=80mm,width=100mm,bbllx=0pt,bblly=0pt,bburx=600pt,bbury=650pt,angle=270}}
{\bf Figure 3.}
{\sl The real component of the third harmonic susceptibility
as a
function of the amplitude of the ac field, computed for an
ac field of period
equal to 25000. The slope of solid lines decreases for
increasing values of $\kappa$,which are the following:
0.667, 0.8, 1.2, 1.333, 1.667, 1.8, 2.333, 2.7, 3.35, 3.7.
Similar scaling property is observed for the imaginary
component of the third harmonic susceptibility and for
higher-order components as well.}
\vspace{5mm}

   {\bf 4. Conclusions}

    When the ac magnetic field does  not penetrate to the
sample center, the magnetic susceptibility is well described
by a
simple scaling relation:
$\chi \sim \xi$, with
$\xi = t^{1/(\kappa +2)}\cdot H_0 ^{\kappa /(\kappa +2)}/D$.
For $\kappa$ values close to $0$,
the overall dependence of  $\chi '' (\chi ')$
closely resembles
the dependence observed in the linear case. The
conductivity,
however, computed from $\chi '' (\chi ')$ data by using the
assumption
that the linear theory holds, will lead to false information
and
yield strongly overestimated values. An experimental
criterion
for detecting nonlinearity would be the observation of the
amplitude dependence or the existence of higher order
harmonics in the ac response. The susceptibility values of a
nonlinear medium are not uniquely defined by a set of two
parameters only, such as $\kappa$ and $\xi$.  For
experimental purposes,
however, treating $\xi$ as a scaling variable offers a
sufficiently
accurate method of testing for nonlinear properties of
materials.

\vspace{5mm}

\centerline{\psfig{figure=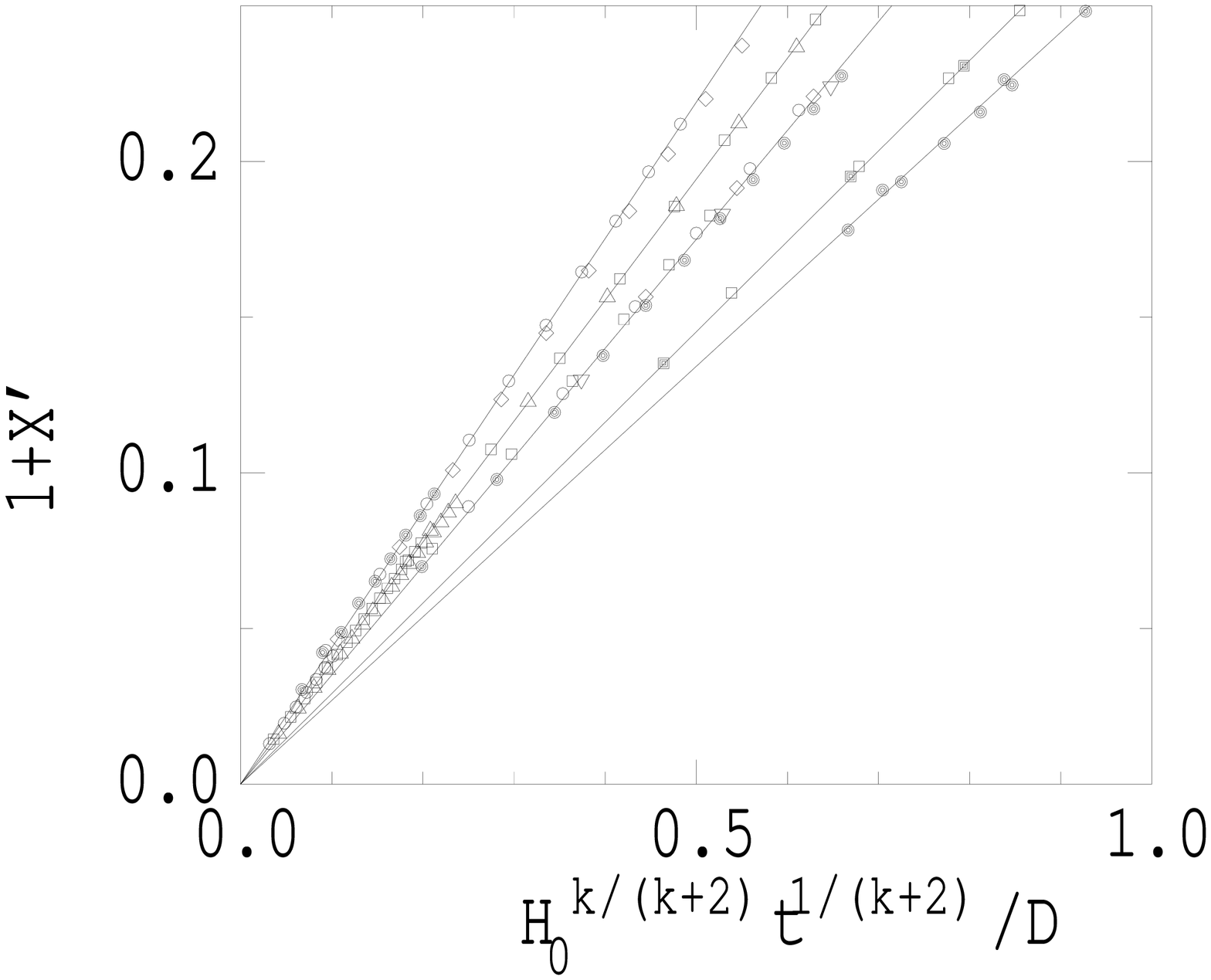,height=80mm,width=100mm,bbllx=0pt,bblly=0pt,bburx=600pt,bbury=650pt,angle=270}}
{\bf Figure 4.}
{\sl  Scaling of the flux penetration $(4\pi \chi ' _1+1)$
by a  function
$H_0 ^{\kappa /(\kappa +2)} \cdot t^{1/(\kappa +2)}/D$,
where $H_0$ gives the field amplitude, $t$ gives the period
and
$D$ gives the sample thickness. Each of the solid lines
passes through the data
corresponding to the following values of the nonlinearity
parameter $\kappa$: 0.667, 1, 2, 3 and 5,  for lines with
the smallest to largest slope. $D$ is equal to 20 or
100 (there is no distinction between the symbols of the data
corresponding to different values of $D$), while $t$ is 2500
($\circ$), 5000 ($\Box$), 6250 ($\Diamond$), 10000
($\Delta$), 12500 ($\nabla$), 25000 ($\bullet$) and 100000
($\Box$).}
\vspace{5mm}

   {\bf Acknowledgements} --- This work has been made
possible due to the
Killam Memorial Postdoctoral Fellowship awarded to Z.K. by
Dalhousie University and a grant from the Natural Sciences
and
Engineering Research Council of Canada.

\renewcommand{\refname}{\hspace{4cm} ---------------------}


\begin{thebibliography}{9}
\bibitem{ref9} The source code is available from the
authors.
\end{thebibliography}
\end{document}